\documentclass[10pt,a4paper,draft,aps,prl,twocolumn,superscriptaddress,nofootinbib]{revtex4}
\usepackage[latin1]{inputenc}
\usepackage{amsmath}
\usepackage{amsfonts}
\usepackage{amssymb}
\usepackage[final]{graphicx}
\usepackage[normalem]{ulem}
\usepackage[usenames, dvipsnames]{color}
\usepackage{soul}
\usepackage{color}

\begin{document}
\title{Geometric Heat Engines Featuring Power that Grows with Efficiency}
\author{O. Raz}
\author{Y. Suba\c{s}\i} 
\affiliation{Department of Chemistry and Biochemistry ,
	University of Maryland, College Park, MD 20742, U.S.A.}
\author{R. Pugatch}
\affiliation{Simons Center for Systems Biology, School of Natural Sciences, Institute for Advanced Study, Princeton, NJ 08540, U.S.A.}
\begin{abstract}
Thermodynamics places a limit on the efficiency of heat engines, but not on their output power or on how the power and efficiency change with the engine's cycle time. In this letter, we develop a geometrical description of the power and efficiency as a function of the cycle time, applicable to an important class of heat engine models. This geometrical description is used to design engine protocols that attain both the maximal power and maximal efficiency at the fast driving limit. Furthermore, using this method we also prove that no protocol can  exactly attain the Carnot efficiency  at non-zero power. 
\end{abstract}	
\maketitle

\paragraph{Introduction}   
Heat engines - machines that exploit temperature differences to extract useful work, are modeled as operating in either a non-equilibrium steady-state, e.g. thermoelectric  \cite{AmJPhys_1991_ThermoElectricEngine,EPL_2009_NanoThermoElectric} or chemical potential  \cite{PRL_2009_universality} driven engine, or as a cyclic engine, where external parameters and temperature are varied periodically in time, e.g. the Carnot, Otto, Stirling and the Diesel cycles \cite{callen_thermodynamics}. Both types of engines are characterized by two main figures of merit: efficiency and power. 

In steady-state heat engines, currents generated by the temperature difference flow against some affinities (thermodynamical forces), e.g. electrical \cite{AmJPhys_1991_ThermoElectricEngine,EPL_2009_NanoThermoElectric,PhysRevE.Two_Thermoelectric_Coupled_Engines} or chemical potentials \cite{PRL_2009_universality}, generating useful work. From a practical standpoint,  the interest in these engines is limited, since the majority of heat engines are better modeled as cyclic engines. One of the main motivations to study steady-state heat engines, however, is the hope that they share universal characteristics with cyclic engines, which are generically more difficult to analyze. An example for such a characteristic behavior is the relationship between power and efficiency: in all steady state heat engines,
the affinity at maximal power does not equal to the affinity at maximal efficiency, unless one of the heat baths is at infinite or zero temperature. Heat engines that attain their maximal power and maximal efficiency (which is either the Carnot efficiency or a lower value) at different working conditions are  here defined as \emph{heat engines with a power-efficiency trade-off}. The power-efficiency trade-off is the subject of many recent studies \cite{PhysRevE.Eff_and_Power_Linear_Response,PRL_2005_VDB_Efficiency,PRL_2009_universality,PRL_2011_TimeRevers_Eff,PRL_2012_Kinesine_efficiency,PRL_2014_FiniteBathes,PhysRevB_Thermoelectric_PowerEff_Tradeoff}.    

Less is known about the efficiency and power of cyclic heat engines, but a lot of research effort has been devoted to understanding them in recent years \cite{PhysRevX_Seifert_2015,VanDenBroeck_PRL_2015,VanDenBroeck_2010_PRL,PRE_2012_efficiency,Siefert2008SimpModel,PhysRevE_CyclicEngineTradeoffs,PhysRevE_CarnotCycle_Finite_Time_Lear_Response,PhysRevLett_CarnotCycle_ProteinFolding}. The operation of a cyclic engine is characterized by a protocol that describes the time dependence of key variables along the cycle --- e.g. piston position and temperature. The set of feasible protocols however, is strongly bounded by a set of engine specific and hence non-generic constraints. 
Maximizing power or efficiency is, therefore, a nontrivial constrained optimization problem. Nevertheless, there is a natural optimization problem in these engines which is both simpler and of practical importance: optimization with respect to overall cycle time. In most cyclic engines, the protocol is determined up to a rescaling of the cycle time. In other words, the overall rate at which the engine operates can be varied. It is thus natural to consider the efficiency and power of a heat engine with a fixed protocol as a function of the cycle time. 

Steady-state heat engines have a characteristic trade-off between power and efficiency as a function of the affinity. Do cyclic heat engines have a corresponding power-efficeincy trade-off as a function of their cycle time? Analytical \cite{Siefert2008SimpModel}, numerical \cite{Tradeoff_numerical_diesel,Tradeoff_numerical_Atkinson} and experimental \cite{NatPhy_2012_ExpEngine_NegPower} results for certain driving protocols seem to suggest that this might be the case: increasing the cycle time increases the efficiency, with the maximal efficiency (which is possibly lower then the Carnot limit as in the Diesel, Miller and Sargent cycles) only attained in the quasi-static limit --- namely at infinitely long cycle time, where the power vanishes. On the other hand, driving the engine faster increases the power at the expense of efficiency, until eventually, at fast enough driving, the dissipation rate becomes significant and causes a decrease in power. Yet, as we demonstrate, this behavior is not universal, and there is no inherent trade-off between power and efficiency as a function of the cycle time, although such a trade-off always exists in cycles that exactly attain the Carnot bound.

Here we analyze a class of cyclic heat engine models, referred to as \emph{geometric heat engines}, which includes the paradigmatic examples of a Brownian particle in a parabolic potential \cite{Siefert2008SimpModel,VanDenBroeck_PRL_2015,PhysRevX_Seifert_2015,NatPhy_2012_ExpEngine_NegPower} and the two-state Markovian engines \cite{NJP_TwoStateEngien}, but is not limited to these models. In this class,  the work and heat can be interpreted as areas in state space (the space of all the possible states of the engine) defined by the periodic trajectory of the engine's state. This interpretation has two consequences: (i) Work and heat can be formulated as  parametrization independent quantities, and (ii) Power can vanish due to the formation of singularities in the engine's trajectory in state space. We then introduce a time re-parametrization that effectively decouples the state space variables. This decoupling, together with an understanding of the aforementioned singularities, enables the design of novel engine driving protocols that avoid power losses at short cycle times. As an example, we construct a protocol whose power and efficiency are both maximized at the infinitely fast driving limit. This proves that cyclic heat engines do not have an inherent trade-off between power and efficiency as a function of their cycle time. Achieving the maximum efficiency at finite power, however, comes with a price: we prove that in this class of models the Carnot limit cannot be attained at non-zero power. Therefore, to avoid the trade-off, the efficiency must be lower then the Carnot limit.  Similar relations between the Carnot limit and power were discussed in \cite{VanDenBroeck_PRL_2015} in the linear response regime , and very recently for arbitrary Markovian dynamics with local detailed balance in \cite{NewArxiv}. 

\paragraph{Model description:}

For simplicity, we focus here on a specific model, and subsequently show that our results are valid for a larger class of models. This model consists of an overdamped Brownian particle confined to one spatial dimension, in a time dependent harmonic potential $V(x,t) =\frac{\Lambda(t)}{2}x^2$, coupled to a heat bath with a time dependent inverse temperature $\beta(t)$. This model was suggested in \cite{Siefert2008SimpModel} and experimentally realized in \cite{NatPhy_2012_ExpEngine_NegPower}.  Both $\Lambda(t)$ and $\beta(t)$ are periodic with a cycle time $\tau$ \footnote{To establish the geometrical picture we assume that both $\Lambda$ and $\beta$ are twice differentiable.}. The engine's protocol, which is a time-parametrized closed curve in the \emph{control space} - the space of external control parameters, is denoted by $\Gamma_t^{\beta,\Lambda} = (\beta(t),\Lambda(t))$ where the subscript $t$  indicates that the protocol is parametrized by the time and the superscripts $\beta,\Lambda$ indicate that this protocol is defined in the $[\beta,\Lambda]$ control space.

    The probability density to find the particle at point $x$ at time $t$, denoted by $p(x,t)$, evolves according to the Fokker-Planck equation,  $$\partial_t p(x,t) = -\frac{1}{2}\partial_x\left[-\partial_x V- \beta^{-1}\partial_x\right]p(x,t),$$ where for simplicity we assumed the mobility to be one half. Under this evolution, an initial Gaussian distribution centered at the origin remains a centered Gaussian with time dependent width and height \cite{Siefert2008SimpModel},  which are related through probability conservation. This Gaussian distribution can be parametrized by a single parameter -- the variance, $w(t) = \langle x^2\rangle$, which evolves in time according to \cite{Siefert2008SimpModel}
\begin{eqnarray}\label{Eq:w_dot_t}
\frac{dw}{dt} = -\Lambda(t) w(t) + \beta^{-1}(t).
\end{eqnarray}
After sufficiently long time, the system reaches a periodic state, $w(t) = w(t + \tau)$. We assume that the engine already relaxed to this periodic state.

To describe the engine's properties when the same protocol is performed at different cycle times, it is useful to consider the protocol in terms of a dimensionless time parameter, $ s  = t/\tau$ in the interval $[0 ,1)$, rather than $0\leq t<\tau$. Defining the protocol as $\Gamma_s^{\beta,\Lambda} = (\beta( s ),\Lambda( s ))$, allows us to treat the cycle time $\tau$ as a parameter, independent of other characteristics of the protocol.  The time evolution of the width $w$, governed by Eq.(\ref{Eq:w_dot_t}), can be written in terms of $ s $ as:
\begin{eqnarray}\label{Eq:w_dot}
\frac{dw}{d s }= \tau\Big(\beta^{-1}(s)-\Lambda(s) w(s)\Big) .
\end{eqnarray}
Note that the solution $w(s)$ of the above equation is explicitly $\tau$ dependent.  

The infinitesimal heat exchange between the system and the bath is given by $dQ = \frac{\Lambda}{2} \frac{dw}{ds} ds$. With this convention, $dQ>0$ implies that heat flows into the system \cite{Siefert2008SimpModel,seifert2012stochastic}. By conservation of energy, the total work extracted in a cycle and the corresponding power are given by 
\begin{eqnarray}
W = \int_0^\tau dQ = \int_0^1 \frac{\Lambda}{2} \frac{dw}{d s }d s  ;&P=\frac{W}{\tau}.
\end{eqnarray}
The equation for the work $W$ has a geometrical interpretation: $W$ is half the oriented area bounded by $\Gamma_s^{w,\Lambda}=(w( s ),\Lambda( s ))$, namely by the curve in the $[w,\Lambda]$ state space.  An important consequence of the geometrical interpretation is that the work is parametrization independent. By this we mean that if some other driving protocol, $\hat{\Gamma}_s^{\beta,\Lambda}$, operating at some cycle time $\hat{\tau}$, happens to trace the same curve in the $[w,\Lambda]$ space as $\Gamma_s^{w,\Lambda}$ but with a different $s$ parametrization, then the extracted work is equal in the two protocols, even though $\Gamma_s^{w,\Lambda}\neq\hat\Gamma_s^{w,\Lambda}$. 

To define the efficiency, we need to quantify the cost of any protocol in terms of the heat extracted from the heat baths during a cycle. This can be done by  accounting for only the sections of the cycle during which heat flows from the bath into the system:
\begin{eqnarray}
Q_{in} = \int_0^1 \frac{\Lambda}{2} \frac{dw}{d s }\Theta\left[\Lambda \frac{dw}{d s }\right]ds ,
\end{eqnarray}     
where $\Theta[\cdot]$ is the Heaviside step function. This integral can be interpreted geometrically as half the area under the sections of $\Gamma_s^{w,\Lambda}$ in which $w$ decreases \footnote{Note that the term $\Theta\left[\Lambda \frac{dw}{d s }\right]$ is not geometrical, since it is not parametrization independent. However, if we consider only re-parameterizations $ s \rightarrow\lambda $ in which $\lambda ( s )$ is a monotonically increasing function, then the geometric interpretation can be applied.}. With these definitions of work $W$ and heat $Q_{in}$, which are consistent with the laws of thermodynamics \cite{Siefert2008SimpModel}, the efficiency is given by $\eta = \frac{W}{Q_{in}}$, and it can be interpreted as the ratio between the two corresponding areas \footnote{This definition is not the only possible definition, see e.g.  \cite{PhysRevX_Seifert_2015}. However the geometric picture associate with it is a notable advantage. Moreover, other definitions of efficiency show the same qualitative behavior.}.

The interpretation of the work and heat as areas is done in the $[w,\Lambda]$ state space , which describes the instantaneous state of the engine, whereas the engine's driving protocol is defined in the $[\beta,\Lambda]$ control space, which is the space of the external control parameters. This makes it difficult to directly relate the protocol to the areas associated with work and heat. However, the protocol can be defined equally well in terms of $\Omega= \left(\beta\Lambda\right)^{-1}$ instead of $\beta$, namely as $\Gamma_s^{\Omega,\Lambda}=(\Omega(s),\Lambda(s))$. Note that $\Omega$ is the width of the Boltzmann distribution for the potential $V=\Lambda\frac{x^2}{2}$. The main advantage of defining the protocol as $\Gamma_s^{\Omega,\Lambda}$ is that in the quasi-static limit, $\tau\rightarrow\infty$, Eq.(\ref{Eq:w_dot}) implies that $w( s )\rightarrow\Omega( s )$ and $\Gamma_s^{w,\Lambda}\rightarrow\Gamma_s^{\Omega,\Lambda}$, therefore we can unify the control space and state space. With decreasing $\tau$, the contour that defines the protocol, $\Gamma_s^{\Omega,\Lambda}$, continuously deforms into the contour $\Gamma_s^{w,\Lambda}$, which has the geometrical interpretation for work and heat. Therefore, we study this deformation next. 
 
\begin{figure}
\centering
\includegraphics[width=\linewidth]{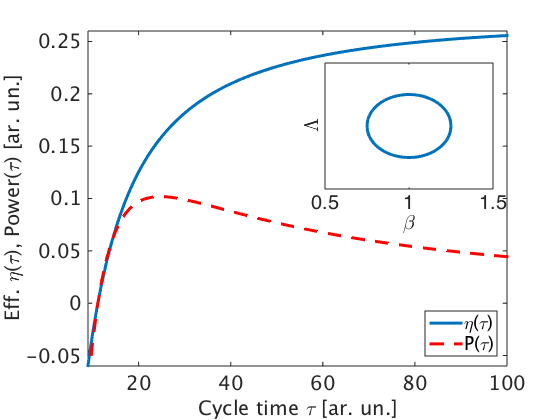}
\includegraphics[width=\linewidth]{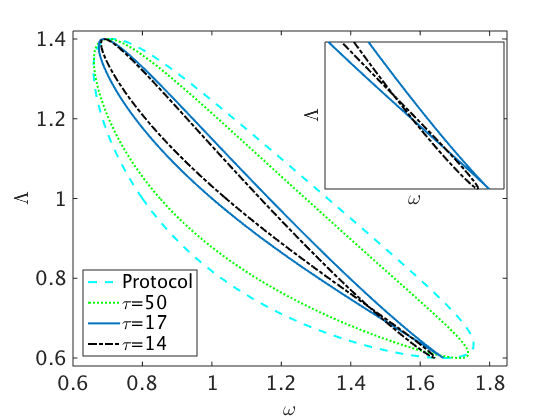}
\caption{{\bf Upper panel:} Power and efficiency as a function of the cycle time $\tau$ for a circular protocol in the $[\beta,\Lambda]$ control space (inset). This protocol attains its maximal power ($\eta_{max}\approx0.26$) at the $\tau\rightarrow\infty$ limit. The Carnot limit for this protocol is $\eta_C = 0.4$. At high driving rates, the power and efficiency change their sign. {\bf Lower panel:} The $[w,\Lambda]$ curves of the above protocol for few values of the cycle time $\tau$. At about $\tau=17$ the curve develops a cusp, which for even smaller $\tau$ evolves into a loop with a negative orientation. The inset shows a blow-up of the cusp and a negatively oriented loop.}
\label{fig:Circular_protocol}
\end{figure}

\paragraph{Finite cycle time:}
Consider first a simple driving protocol, in which $\Gamma_t^{\beta,\Lambda}$ traces a circle at a uniform rate in control space  as shown in the upper inset of Fig(\ref{fig:Circular_protocol}). The upper panel shows the efficiency and power as a function of the cycle time $\tau$. As can be seen, this protocol has the typical behavior described above: the efficiency is a monotonically increasing function of $\tau$, attaining its maximum as  $\tau\rightarrow\infty$. The power is zero where the efficiency is maximal, and attains its maximum at about $\tau=25$. Below about $\tau=11$, both the power and efficiency  become negative, since the work changes its sign.  For the work to become negative, the area bounded by $\Gamma_s^{w,\Lambda}$ must changes its orientation.  How can this comes about? 

There are only two generic ways in which the area orientation of a closed curve in 2D can change through continuous deformations: a cusp singularity or a self-tangent singularity \cite{Book_hilton1920plane}. We next analyze the formation of the cusp singularity, whereas the formation of the self-tangent singularity is analyzed in the SI. A cusp in $\Gamma_s^{w,\Lambda}$ emerges when both $w( s )$ and $\Lambda( s )$ have an extreme point at the same value of $s$ \cite{Book_hilton1920plane}.  In other words, if varying $\tau$ causes the value of $s^*$ at which $\frac{dw}{ds}( s^* )=0$ to pass through  $s^{**}$ for which $\frac{d\Lambda}{ds}(s^{**})=0$, then a cusp is formed when $s^* = s^{**}$, and developed into a loop with an inverted orientation. This loop decreases the power and efficiency, enabling the work to vanish at some non-zero $\tau$. In the specific example described in Fig.(\ref{fig:Circular_protocol}), a cusp singularity is generated at $\tau = 17$. Below this $\tau$, the cusp evolves into a negatively oriented loop, reducing the power and efficiency. 

Realizing that the deterioration of the power can be related to the formation of negatively oriented areas immediately raises the question: can a protocol, in which negatively oriented areas never occur, be designed?  To avoid the singularities,  the locations $s^*_i (\tau)$ of the extreme points of $w( s )$, where the index $i$ label the different extreme points, must be considered. However, even if we know $s^{*}_i(\tau)$, manipulating the protocol to avoid the singularities is challenging, since varying either $\Lambda(s)$ or $\Omega(s)$ varies $s^*_i(\tau)$ as well. To simplify the analysis, we take advantage of the fact that areas are parametrization invariant. Instead of looking on the actual parameterization, we consider the following re-parametrized quantities
\begin{eqnarray}
\bar{w}(s) = w\Big(\lambda(s)\Big),\;\;\bar{\Lambda}(s) = \Lambda\Big(\lambda(s)\Big), \;\; \bar{\Omega}(s) = \Omega\Big(\lambda(s)\Big)
\end{eqnarray} 
where the re-parametrization $\lambda(s)$ is given by
\begin{eqnarray}\label{Eq:lambda_s}
\lambda(s)= \frac{\int_{0}^{ s }\bar\Lambda(x)^{-1}dx}{\int_{0}^{1 }\bar\Lambda(x)^{-1}dx}.
\end{eqnarray}  
Note that $\frac{d\lambda}{d s }>0$, i.e. $\lambda(s)$ is monotonically increasing with $s$, and moreover $\lambda(0)=0$ and $\lambda(1)=1$. Also note that the transformation is given in terms of $\bar{\Lambda}(s)$ rather then $\Lambda(s)$.  This turns out to be useful in what follows.  Although $\bar w(s)\neq w(s)$ and $\bar{\Lambda}(s)\neq \Lambda(s)$, their corresponding curves, $\Gamma_s^{w,\Lambda}$ and $\Gamma_s^{\bar w, \bar \Lambda}$, trace \emph{the same contour} in the engine's state space, $[w,\Lambda]$, hence they have the same work and efficiency.   In this re-parameterization,
\begin{eqnarray}\label{Eq:w_vs_r}
\frac{d\bar w}{ds} = \bar\tau\left(\bar\Omega(s)-\bar w(s)\right)
\end{eqnarray}
where $\bar{\tau} = \frac{\tau}{\int_0^1\bar\Lambda(x)^{-1}dx}$.
Although $\Gamma_s^{w,\Lambda}\neq\Gamma_{\lambda(s)}^{w,\Lambda}=\Gamma_s^{\bar w,\bar\Lambda}$, the two curves are different parameterizations of the same contour in the $[w,\Lambda]$ state space, and hence they enclose the same area and represent heat engines with the same values of heat and work. The main advantage of this re-parametrization is that the equation for $\frac{d\bar w}{ds}$ (Eq. \ref{Eq:w_vs_r}) is independent of the potential width $\bar \Lambda$. Therefore, using the $\lambda$ parametrization to define the protocol, $\bar\Lambda(s)$ can be manipulated without affecting $\bar w(s)$, but  at the price of rescaling $\tau$. 

\paragraph{The $\tau$ dependence of $\bar w(s)$'s extreme points.} 

\begin{figure}
\centering
\includegraphics[width=\linewidth]{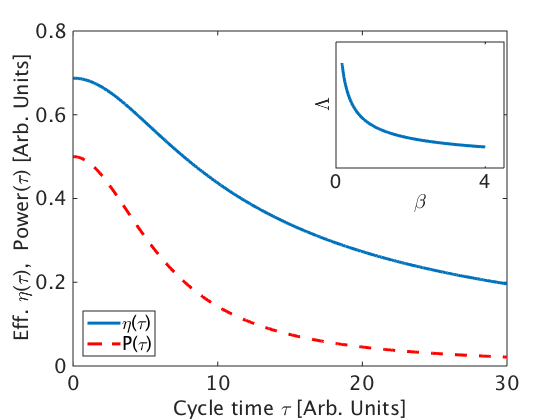}
\includegraphics[width=\linewidth]{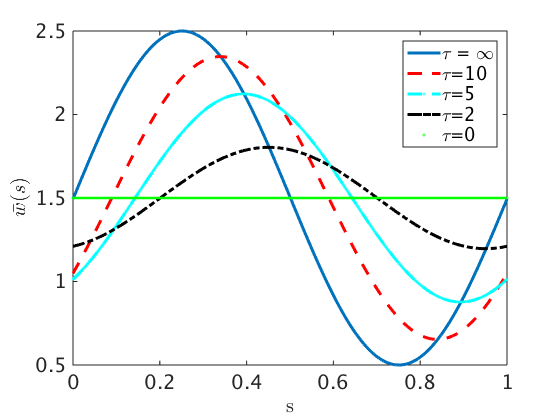}
\caption{{\bf Upper panel:} Power and efficiency as a function of the cycle time $\tau$. The protocol is given by $\bar \Omega(s) = \bar\Lambda(s) =1.5 + \sin(2\pi s)$. The inset shows the protocol in the $[\beta,\Lambda]$ control space. The protocol traverses the line back and forth, without covering any area. {\bf Lower panel:} $\bar w(s)$ for various values of $\tau$. When $\tau=\infty$, $\bar w(s)=\bar \Omega(s)$. As can be seen, the maximum of $\bar w(s)$ ``slides'' down the negative slope when $\tau$ decreases, and the minimum if $\bar{\omega}(s)$ ``slides up'' on the positive slope. Overall, $\bar w(s)$ shifts to the right and its amplitude decreases with decreasing $\tau$.} 
\label{fig:Resub_fig_2_Eff_Power_Graph}
\end{figure}

To avoid cusp generation, we next explain how the extreme points of $\bar w(s)$ change as a function of $\bar\tau$. Let us denote by $s^*$ any  $s$ at which $\bar w(s)$ has an extreme point, namely $\frac{d\bar w}{ds}(s^*) =0$. By taking the derivative of Eq.(\ref{Eq:w_vs_r})  and using $\frac{d\bar w}{ds}(s^*)=0$ it follows that 
\begin{eqnarray}
\bar w(s^*) =\bar\Omega(s^*);& &
\frac{d^2\bar w}{ds^2}(s^*) = \bar\tau\frac{d\bar\Omega}{ds}(s^*).
\end{eqnarray}
These two equations can be interpreted as follows: In the quasi-static limit, $\hat{\tau}\rightarrow\infty$, $\bar w(s)=\bar \Omega(s)$ everywhere, including the extreme points. Decreasing $\bar\tau$, the maximal points of $\bar w(s)$, for which $\frac{d^2\bar w}{ds^2}(s^*)<0$, ``slide'' along the $\bar\Omega(s)$ curve down and to the right (where the slope of $\bar\Omega(s)$ is negative), and similarly, the minimal points of $\bar w(s)$ (where $\frac{d^2\bar w}{ds^2}(s^*)>0$) slides up and to the right along the positive slope of $\bar \Omega(s)$. Physically, this means that decreasing $\bar\tau$ results in a flattening of $\bar w(s)$, with an increasing phase-lag between $\bar w(s)$ and $\bar \Omega(s)$. Example for this behavior can be seen in the lower panel of Fig.(\ref{fig:Resub_fig_2_Eff_Power_Graph}). 

At this point one might worry about a different effect that have the potential to decrease the power with decreasing $\bar\tau$: the range covered by $\bar w(s)$, namely the range between its maximum and minimum, decreases with decreasing $\bar\tau$, eventually shrinking the area. This might cause the power to decrease, even without the phase-lag, which is manifested e.g. in the generation of negatively oriented loops. In the SI we show that for the specific model considered here, this is not the case.

\paragraph{Designing a protocol without a power-efficiency trade-off.}

To avoid the power deterioration, we should design a protocol that does not form singularities. This can always  be done by choosing $\bar \Omega(s)=c_1\bar\Lambda(s) + c_2$ for some constants $c_1$ and $c_2$, and $\bar\Lambda(s)$ that has a single minimum and maximum. In such a protocol, the extreme points of $\bar w(s)$ coincide with those of $\bar\Lambda(s)$ only in the $\bar\tau\rightarrow\infty$ limit. As discussed in the SI, such a protocol is also assured not to have a self tangent singularity for any finite $\bar\tau$. In the $\bar\tau\rightarrow\infty$ limit, $\Gamma_s^{\bar w,\bar\Lambda}$ bounds no area and the work and efficiency are zero. When $\bar\tau$ decreases, the phase lag between $\bar\Lambda(s)$ and $\bar w(s)$ `inflates' the area.  For demonstration, we show in Fig.(\ref{fig:Resub_fig_2_Eff_Power_Graph}) the example $\bar\Lambda(s) = \bar\Omega(s) = 1.5 + \sin(2\pi s)$, for which $\bar w(s)$ can be solved analytically. As shown in the figure, when $\bar\tau$ decreases, $\bar w(s)$ shifts to the right and its amplitude decreases. This protocol indeed does not have any trade-off between power and efficiency, and in fact both the maximal power and maximal efficiency are attained asymptotically at the fast driving limit. 

To implement this protocol in an experimental realization as in \cite{NatPhy_2012_ExpEngine_NegPower}, it is not useful to specify it as  $\bar\Lambda(s)$ and $\bar\Omega(s)$, since in the laboratory the protocol $\Lambda( s )$ and $\beta( s )$ should be performed. However, as $\lambda(s)$ is given in terms of $\bar{\Lambda}$ (Eq. \ref{Eq:lambda_s}), it can be calculated from the designed protocol, inverted into $\lambda^{-1}(s)$ such that $\lambda^{-1}\left(\lambda(s)\right)=s$, and then $\Lambda(s) = \bar{\Lambda}\left(\lambda^{-1}(s)\right)$ and $\Omega(s) = \bar{\Omega}\left(\lambda^{-1}(s)\right) $.

The resulting protocol in the $[\beta,\Lambda]$ space is shown in the inset of Fig.(\ref{fig:Resub_fig_2_Eff_Power_Graph}). Note that the maximal efficiency in this protocol is only $\eta_{max}=0.687$ (attained in the limit $\tau\rightarrow 0$), compared to the Carnot efficiency calculated from the minimal and maximal temperatures which is $\eta_{Carnot}= 0.96$. This is expected, since in this protocol the engine exchanges heat with many heat baths at intermediate temperatures, hence it cannot attain the Carnot efficiency computed with the extreme temperatures. However, \emph{this protocol does not suffer from a tradeoff between power and efficiency.}   

With the tools developed above, it is natural to ask: is it possible to design a protocol that achieves the Carnot limit but has non-zero power, namely operate at a finite cycle time $\tau$? As we show in the SI, the answer is no. This is proven by showing that for any piecewise continuous protocol $\Gamma_s^{\beta,\Lambda}$ attaining the Carnot efficiency at some finite cycle time $\tau$, it is possible to slightly deform the protocol into a new protocol $\tilde\Gamma_s^{\beta,\Lambda}$ which has the same $\beta(s)$ (hence the same Carnot bound) but different $\Lambda(s)$, such that the efficiency of the new protocol $\tilde\Gamma_s^{\beta,\Lambda}$ is strictly larger then the efficiency of $\Gamma_s^{\beta,\Lambda}$, hence larger then the Carnot limit. This would violate the second law. 

\paragraph{Applicability to other models.}
The above analysis was made possible due to three properties of the model: (i) The interpretation of the work and incoming heat as areas; (ii) The 2D nature of the state space, which enabled simple characterization of all the possible ways in which the area can change its orientation, and (iii) The ability to separately control the work parameter $\bar\Lambda(s)$ without influencing the other state parameter, $\bar w(s)$. This required changing the time parametrization from $s$ to $\lambda(s)$. What class of models share these properties? The first property is common in a wide range of models, including any Markovian finite state system and any Brownian engine in which both the time dependent potential $V(x,t)$ and the periodic solution for the probability density $p(x,t)$ can be described by a finite number of parameters at each given time $t$. The second property is not a true limitation, however, at higher dimensions the analysis of the possible ways in which the area changes its sign is significantly complicated. We do not know what is the exact class of models in which the last condition is possible, however, as shown in the SI, in addition to the example given above, this class contains also any 2-levels Markovian model.    
\paragraph{Acknowledgments}
We would like to thank C. Jarzynski for useful discussions and Y. Sagi, S. Rahav, S. Kotler, O. Hirshberg, V. Alexandrov, J. Hopfield for discussion and comments. O.R. acknowledges the financial support from the James S. McDonnell foundation. R.P. would like to thank  S. Christen and A. Levine for their help and support. Financial support from Fullbright foundation, Eric and Wendy Schmidt fund and the Janssen Fellowship are greatfully acknowledged. Y.S. acknowledge financial support from  the U.S. Army Research Office under contract number W911NF-13-1-0390.


\appendix
\newpage
\begin{widetext}
	\centering
	\LARGE Supplementary Information\\[1.5em]
\end{widetext}
This supplementary information has 5 sections: In the first section, we consider the self-tangent singularity; In the second section, we discuss the transformation $\lambda(s)$; In the third section  we show that any decrease in the power at low $\tau$ comes only from the phase between $\bar w$ and $\bar{\Omega}$ and not from the amplitude of $\bar w(s)$; In the fourth section, we show additional important example for geometrical heat engine; In the last section, we show that no protocol in these engines can attain the Carnot efficiency with non-zero power.   
\section{Formation of negatively oriented section through the ``self-tangent'' mechanism.}
\begin{figure}
	\centering
	\includegraphics[width=0.7\linewidth]{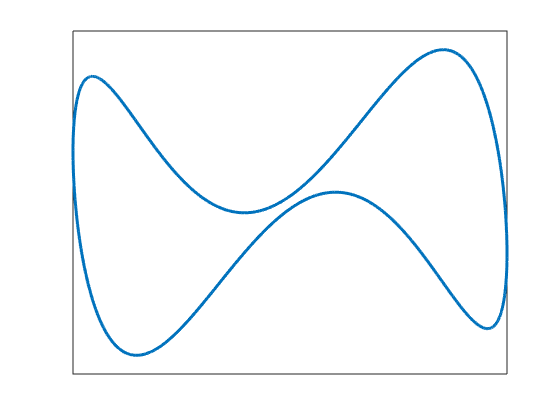}\\
	\includegraphics[width=0.7\linewidth]{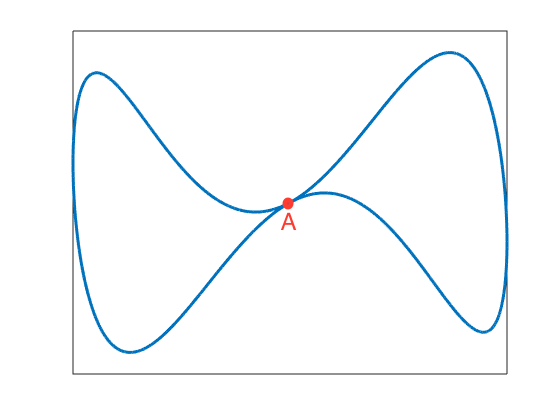}\\
	\includegraphics[width=0.7\linewidth]{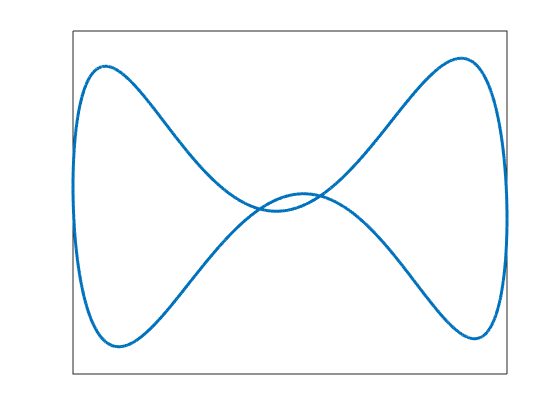}
	\caption{Upper panel - a curve which does not intersect itself. The orientation of the area bounded by the curve has a single sign. Lower panel - a self intersecting curve. The left most and right most areas have the same orientation, but the middle section has an opposite orientation. Middle panel - for the curve to develop a section with opposite orientation through self intersection, it must pass through a self-tangent point marked by $A$.}
	\label{fig:Supinfo_Kiss_1}
\end{figure}

In the main text we discussed the generation of negatively oriented loops in $\Gamma_{s}^{\bar w,\bar \Lambda}$ through a cusp singularity. In this section we discuss the second mechanism for generating negatively oriented area --- the formation of a self tangent. In this process, one section of the curve intersects a different section - see the lower panel in Fig.(\ref{fig:Supinfo_Kiss_1}). 
Let us write the necessary and sufficient conditions for 
$\Gamma_s^{\bar w,\bar\Lambda}$ to have a point at which it is tangent to itself for some $\hat \tau$ (e.g. point $A$ in the middle panel in Fig.\ref{fig:Supinfo_Kiss_1}):
\begin{eqnarray}
\bar w( s _1) &=&\bar w( s _2)\\ 
\bar\Lambda( s _1) &=& \bar\Lambda( s _2)\\
\frac{ d\bar\Lambda/ds(s_1)}{d\bar\Lambda/ds(s _2)} &=& \frac{d\bar w/ds(s_1)}{d\bar w/ds(s_2)} 
\end{eqnarray}
The first two equations assure that the curve intersects itself, and the last equation assures that at the intersection point the curve is tangent to itself. Using the equation for $\frac{d\bar w}{ds}$ given in the main text, the right hand side of the last equation can be expressed as
\begin{eqnarray}
\frac{d\bar w/ds(s_1)}{d\bar w/ds(s_2)} = \frac{\bar\Omega(s_1)-\bar w(s_1)}{\bar\Omega(s_2)-\bar w(s_2)}  
\end{eqnarray}
However, at the tangent point $\bar w( s_1) =\bar  w(s_2)$. We can therefore write
\begin{eqnarray}\label{Eq:Edot_ratio}
\frac{ d\bar\Lambda/ds(s_1)}{d\bar\Lambda/ds(s_2)}  = \frac{\bar\Omega(s_1)-\bar w(s_1)}{\bar\Omega(s_2)-\bar w(s_1)}
\end{eqnarray}
Moreover, as discussed in the main text, the range of values that $\bar w(s)$ can have is bounded by the extreme values of $\bar \Omega(s)$. Thus, given $\bar\Omega(s)$, it can be verified for a proposed protocol $\bar\Lambda(s)$ that the  equation above does not hold for any  $ s _1$ and $s_2$ for which  $\bar\Lambda( s _1) = \bar\Lambda( s _2)$ and  any $\bar w$ in the range 
\begin{eqnarray}
\min_ s \{\bar\Omega( s )\}\leq \bar w\leq\max_s \{\bar \Omega( s )\}.
\end{eqnarray}
Note that if $\bar\Lambda(s) = c_1\bar\Omega(s) + c_2$  for some constants $c_1$ and $c_2$, as in the example in the main text, and furthermore $\bar\Lambda$ undergoes only a single oscillation in each period, then Eq.(\ref{Eq:Edot_ratio}) never holds for any finite $\bar\tau$.

\section{Transforming from ${\Lambda(s)}$ to $\bar\Lambda(s)$ and back}

In this section we explain  the transformation from ${\Lambda(s)}$ to $\bar\Lambda(s)$. We start with the equation:
\begin{eqnarray}
\frac{dw}{ds} = \tau\Lambda(s)\Big(\Omega(s)-w(s)\Big)
\end{eqnarray}
For any re-parametrization $\lambda(s)$, the equation for $\bar{w}(s) = w\big(\lambda(s)\big)$ is:
\begin{eqnarray}
\frac{d\bar w (s)}{ds} &=& \frac{d w(\lambda)}{d\lambda}\frac{d\lambda}{ds}\nonumber\\
 &=&\tau\Lambda\Big(\lambda(s)\Big)\left(\Omega\Big(\lambda(s)\Big)-w\Big(\lambda(s)\Big)\right)\frac{d\lambda}{ds}\nonumber\\
 &=&\tau\bar\Lambda(s)\left(\bar\Omega(s)-\bar w(s)\right)\frac{d\lambda}{ds}
\end{eqnarray}
where  $\bar{\Lambda}(s) = \Lambda\big(\lambda(s)\big)$ and $\bar{\Omega} = \Omega\big(\lambda(s)\big)$, as in the main text. Choosing $\lambda(s)$ such that
\begin{eqnarray}
\frac{d\lambda}{ds}=c\bar\Lambda(s)^{-1}
\end{eqnarray} 
for some constant $c$, would cancel the $\Lambda$ dependence in the equation for $\frac{d\bar w}{ds}$. Therefore, we choose 
\begin{eqnarray}
\lambda(s) = \frac{\int_0^s \left(\bar{\Lambda}(x)\right)^{-1}dx}{\int_0^1 \left(\bar{\Lambda}(x)\right)^{-1}dx}
\end{eqnarray}
This specific choice of $c$ makes $\lambda(1)=1$, namely when $s$ varies from $0$ to $1$ both $\bar{\Lambda}(s)$, $\bar{w}(s)$ and $\bar{\Omega}(s)$ complete a period, and hence $\Gamma_s^{\bar w,\bar \Lambda}$ traces exactly the same curve as $\Gamma_s^{w,\Lambda}$. The price for this choice is rescaling of $\tau$: the equation for $\frac{d\bar w}{ds}$ becomes:
 \begin{eqnarray}
 \frac{d\bar w (s)}{ds} &=& \hat{\tau}\left(\bar\Omega(s)-\bar w(s)\right)
 \end{eqnarray}
where 
\begin{eqnarray}
\hat{\tau} = \frac{\tau}{\int_0^1 \left(\bar{\Lambda}(x)\right)^{-1}dx}
\end{eqnarray}

\section{Fourier amplitudes of the power are monotonic with the cycle time $\tau$} 

To show that in the Brownian heat engine model discussed in the main text the phase difference between $\bar w$ and $\bar\Lambda$ is the only reason for the decrease in the power, let us first express both $\bar w(s)$ and $\bar \Lambda(s)$ as a Fourier series: $\bar \Lambda(s) = \sum_{k} \tilde{\Lambda}_k \sin(2\pi k s + \phi^{\Lambda}_k)$ and $\bar w(s) = \sum_{k} \tilde{w}_k \sin(2\pi k s + \phi_k^w)$. In the Fourier basis, the power can be expressed as
\begin{eqnarray}
P = \frac{1}{2\tau}\sum_k 2\pi k \tilde{\Lambda}_k\tilde w_k \sin(\phi^{\Lambda}_k - \phi^w_k).
\end{eqnarray}
From this equation it is evident that the power depends on $\tau$ in three distinct ways: (i) The \emph{phase difference} between $\bar \Lambda$ and $\bar w$. Note that the `drift' of $\bar w(s)$ to the left, discussed in the main text, changes exactly this phase difference; (ii) Besides the phases $\phi_k^w$, there is additional $\tau$ dependence in the term $\frac{\tilde w_k}{\tau}$.  The last factor is (iii) the \emph{spectral overlap} between $\bar\Lambda$ and $\bar w$ -- even if $\tilde\Lambda_k$ is very large for some given $k$, it will not contribute to the power if $\tilde w_k=0$, namely the spectral overlap between them is zero. 

The above three factors are quite general and follows from the definition of the power. Let us next consider the specific model described in the main text. First, we note that in this case the transformation between $\bar\Omega$ and $\bar w$ is linear and time invariant, therefore a frequency shift is impossible and the spectral overlap does not changes with $\tau$. Moreover, the amplitudes $\frac{\tilde w_k}{\tau}$ can be exactly expressed by taking the Fourier series of Eq.(\ref{Eq:w_vs_r}): 
\begin{eqnarray}
\frac{\tilde w_k}{\tau} &=& 
\frac{\tilde\Omega_k}{{4\pi^2k^2+\bar\tau^2}}\\
\phi_k^{w} &=& \phi_k^{\Omega} + \tan^{-1}\left(\frac{2\pi k}{\tau}\right)
\end{eqnarray} 
Note that all the amplitudes $\frac{\tilde w_k}{\tau}$ are monotonically decreasing as a function of $\tau$. Thus, decreasing $\tau$ necessarily increases the amplitudes of all the Fourier coefficients, hence the changes in amplitudes (which shrinks the width in the $w$ direction) cannot explain the power decreasing. The only reason that the power can decreases with decreasing $\tau$ is the phase difference between $\bar w$ and $\bar\Lambda$, which deforms the shape of the curve in the $[w,\Lambda]$ state space.

\section{Markovian Two Levels system }

In the main text we analyzed the Brownian particle in a harmonic potential model for a heat engine. Another simple model for heat engines which is commonly used \cite{NJP_TwoStateEngien} is the Markovian two level system. This engine is made of two states (commonly referred to as ``levels'') in which the system can be at any moment. The energy difference between these levels is denoted by $E$ and it is periodic in time. For simplicity we assume that the lower level's energy is fixed in time at $E=0$, and the upper level's energy is time dependent. The probability to be in the $i$'s level, denoted by $P_i$, varies according to the master equation, $$\dot P_i = \sum_j R_{ij}P_j.$$
In the master equation, $R$ is the rate transition matrix, which can be parametrized by
\begin{eqnarray}\label{Eq:Arrhenius}
R(t) = f(E,\beta)\left(\begin{array}{cc}
-1& e^{\beta(t)E(t)} \\ 
1& -e^{\beta(t)E(t)}
\end{array} \right),
\end{eqnarray}
where $f(E,\beta)$ can be an arbitrary function. This parametrization is physically motivated by  the generalized Arrhenius form . For a two level system, conservation of probability implies $P_2 = 1-P_1$, hence we denote the probability to be in the upper state simply as $P$. The probability to be in the lower level is hence $1-P$. Substituting Eq.(\ref{Eq:Arrhenius}) in the master equation gives 
\begin{eqnarray}
\frac{dP}{dt} &=& f(E,\beta)\left(-(1-P) + e^{\beta(t) E(t)}P\right)\nonumber\\
&=& Pf(E,\beta)\left(1 + e^{\beta(t) E(t)}\right) - f(E,\beta).
\end{eqnarray}
The infinitesimal heat flow from the bath to the system is given by
\begin{eqnarray}
dQ = EdP
\end{eqnarray}
and the work and efficiency can be interpreted, as in the main text, as areas in the $[P,E]$ state space.
Following the same steps explained in the main text, we first change to a dimensionless time units, $s=t/\tau$, to get 
\begin{eqnarray}
\frac{dP}{ds}  = \tau Pf(E,\beta)\left(1 + e^{\beta(t) E(t)}\right) - \tau f(E,\beta).
\end{eqnarray}
In this case, as in the model described in the main text, we cannot vary any of the two control parameters $E(s)$ or $\beta(s)$ without affecting the periodic solution of $P(s)$. We therefore change the time parametrization to
\begin{eqnarray}
\lambda(s) = \frac{\int_0^s f(E(s),\beta(s))^{-1}\left(1 + e^{\beta(s) E(s)}\right)^{-1} ds}{\int_0^1 f(E(s),\beta(s))^{-1}\left(1 + e^{\beta(s) E(s)}\right)^{-1} ds}
\end{eqnarray}  
The equation for $\bar P = P(\lambda(s))$ is given by
\begin{eqnarray}
\frac{d\bar P}{ds}&=&\frac{d P}{ds}\frac{d\lambda}{ds} \nonumber\\
&=& \bar\tau \left[\bar P - \frac{1}{1+e^{\bar \beta \bar E}}\right] = \hat \tau \left(\bar P(s)-\bar \pi(s)\right)
\end{eqnarray} 
where we defined
\begin{eqnarray}
\bar\tau &=& \tau\int_0^1 f(E(s),\beta(s))^{-1}\left(1 + e^{\beta(s) E(s)}\right)^{-1} ds\\\
\bar \pi(s)&=&\frac{1}{1+e^{\bar\beta(s) \bar E(s)}}
\end{eqnarray}
The above equation is exactly analogous to Eq.(6) in the main text, where $[P,E]$ is the state space and $[\pi,E]$ is the new control space. Together with the interpretation of the work and efficiency as areas, the rest of the results can be applied equally well to this model.

\section{No Carnot Efficiency at Finite Time}
\subsubsection{Carnot Efficiency Means Two Temperatures}
To prove that no protocol in the Brownian particle in harmonics potential heat engine model can attain exactly the Carnot efficiency with non-zero power, namely at finite time, we first show that if a protocol achieve exactly the Carnot efficiency, then in a complete cycle heat is exchanged with only two baths: the coldest and hottest one. In other words, while it might be possible that in a cycle with the Carnot efficiency heat is exchanged with many baths at various temperatures, the net heat exchanged in a complete cycle is zero for all the baths, except the one at the hottest and coldest temperatures.  To show this, consider a heat engine which interacts with three heat baths, at temperatures $T_c<T_m<T_h$. Let us denote the total heat exchange with these baths by $Q_c<0$, $Q_m$ (which might be either positive or negative), and $Q_h>0$ respectively. The extracted work is given by $W = Q_h + Q_m + Q_c$. First, consider the case in which $Q_m>0$.  The efficiency in this case is given by 
\begin{eqnarray}
\eta = \frac{Q_c+Q_m+Q_h}{Q_m + Q_h} = 1 + \frac{T_c\Delta S_c}{T_m\Delta S_m + T_h\Delta S_h}.
\end{eqnarray}   
But the total entropy production in a cycle is non-negative. As there is no net entropy production in the system itself after completing a cycle when it is in the periodic state, the total entropy production is given by $\Delta S_c + \Delta S_m + \Delta S_h\geq 0$, thus
\begin{eqnarray}
\eta &\leq&  1 - \frac{T_c\left(\Delta S_m + \Delta S_h\right)}{T_m\Delta S_m + T_h\Delta S_h}\nonumber\\ 
&=&1-\frac{T_c}{T_h - (T_h - T_m)\frac{\Delta S_m}{\Delta S_m + \Delta S_h} }
\end{eqnarray}   
Since $T_h>T_m$ and both $\Delta S_m$ and $\Delta S_h$ are positive, the efficiency is necessarily smaller then the Carnot efficiency.

Similar argument follows for $Q_m<0$: in this case 
\begin{eqnarray}
\eta = \frac{Q_c+Q_m+Q_h}{ Q_h} = 1 + \frac{T_c\Delta S_c + T_m \Delta S_m}{ T_h\Delta S_h}
\end{eqnarray}   
but $\Delta S_c + \Delta S_m + \Delta S_h\geq 0$, thus
\begin{eqnarray}
\eta &\leq&  1 + \frac{-T_c\left(\Delta S_m+\Delta S_h\right) + T_m \Delta S_m}{ T_h\Delta S_h} \nonumber\\
&=& 1 - \frac{T_c}{T_h}+ \frac{ (T_m-T_c) \Delta S_m}{ T_h\Delta S_h} 
\end{eqnarray}   
Since $T_m>T_c$ and  $\Delta S_m<0$, the efficiency is necessarily smaller then the Carnot efficiency.

\subsubsection{Proof for the ``no Carnot efficiency at finite time'' claim}

To show that no piecewise continuous protocol can attain exactly the Carnot efficiency in this model, we will next show by contradiction that if such a protocol exist, then there exist another protocol, with an even larger efficiency. This would clearly violate the second law (which is built into the model), since the efficiency of this other protocol is larger then the Carnot efficiency. 

Assume that there exist a protocol, $\Gamma_s^{\beta,\Lambda}$ that attains exactly the Carnot efficiency with non-zero power at finite time $\tau$. Then, by the argument given above, there is only total heat exchange with two baths, hence $$\eta = \frac{Q_h+Q_c}{Q_h}$$ where $Q_h>0$ and $Q_c<0$ for it to be an engine. Moreover, it must interact for some finite time with the coldest bath, otherwise it will not exchange heat with the coldest bath or with any bath other then the hottest bath, so by the second law its power must be zero. Let us denote one of the intervals in which the engine is in contact with the coldest bath by $(s_1,s_2)$.

In what follows, we shows that it is possible to deform only the part of the protocol in which it touches the cold bath, namely in the interval $(s_1,s_2)$, and increase $Q_c$ (and hence the efficiency) without changing the temperature or the rest of the protocol. To this end we use
\begin{eqnarray}
\frac{\dot w}{\tau} +\Lambda w = \beta^{-1}
\end{eqnarray}
where for simplicity we use the notation $\dot w = \frac{dw}{ds}$. The protocol deformation, denoted by $\delta \Lambda$,  changes also $w$ by $\delta w$, but must not change $\beta^{-1}$, to avoid changing the Carnot efficiency. Therefore, the deformation must be constrained by:
\begin{eqnarray}
\frac{\delta\dot w}{\tau} +\delta(\Lambda w) = \delta (\beta^{-1})=0
\end{eqnarray}
Additional restriction on the deformation comes from the fact that $w$ is a  solution of an o.d.e., thus it must be continues (but not necessarily differentiable), namely $\delta w(s_1)=\delta w(s_2)=0$.

In principle, the deformation is defined by $\delta\Lambda(s)$, and the deformation $\delta w$ can be calculated from the equation above. However, the constraint $\delta w(s_1)=\delta w(s_2)=0$ makes it simpler to define the deformation in terms of  $\delta w$, and solve for $\delta \Lambda$. Let us choose the variation fo the form $\delta w = \epsilon F$ for some function $F$ (and infinitesimal $\epsilon$). Then from the above equation:
\begin{eqnarray}
\frac{\delta\dot w}{\tau} =-\delta(\Lambda w) = -w\delta\Lambda + \epsilon\Lambda F
\end{eqnarray}
But on the other hand 
\begin{eqnarray}
\frac{\delta\dot w}{\tau} =\frac{\epsilon\dot F}{\tau} 
\end{eqnarray}
Equating these two equation and solving for $\Lambda$, one finds that
\begin{eqnarray}\label{Eq:Delta_Lambda}
\delta\Lambda = \epsilon\frac{\Lambda F - \tau^{-1}\dot F}{w}
\end{eqnarray}

Next, let us consider the change in the heat exchange with the cold bath: $$Q_c = \int_{s_1}^{s_2}\Lambda \dot w ds$$ hence
\begin{eqnarray}
\delta Q_c =\int_{s_1}^{s_2}(\delta\Lambda \dot w + \Lambda \delta \dot w) ds 
\end{eqnarray}
Using the results above we can write:
\begin{eqnarray}
\delta Q_c &=& \epsilon\int_{s_1}^{s_2}\left( \dot F\left( \Lambda-\frac{\dot w}{w\tau}  \right) +F\frac{ \Lambda \dot w}{ w}   \right) ds
\end{eqnarray}
Can we choose a deformation $F$ that will make $\delta Q_c>0$, hence increasing the efficiency? To do this we need to choose $F(s)$ such that the above integral is positive, and such that $F(s_1) = 0 $ and $F(s_2)=0$. This can be done as follows: We choose $F(s)$ to be zero everywhere, except for a very small section, of length $\gamma$, somewhere at a location between $s_1$ and $s_2$ such that in this interval $\Lambda(s)$ and $\dot w$ are both continuous. $\gamma$ is chosen to be short enough such that in this interval both $\Lambda-\frac{\dot w}{w\tau}  $ and $\frac{ \Lambda \dot w}{ w}$ are constant to a good approximation, and such that $\frac{\tau \Lambda \dot w}{w}<0$: this is always possible since $w>0$ by definition (it is a variance), and there must be a place where $\dot w<0$ since in a heat engine $Q_c = \int \Lambda \dot wds<0$, and $\Lambda>0$. Let us denote the starting point of the interval by $s_3$. In this interval we choose $$F = -\gamma^{-1}\left(\cos\left(\frac{2\pi(s-s_3)}{\gamma}\right) + 1\right)$$ Note that $F=0$ in both the end points of the interval, and its derivatives at the end points are zero, so $F$ is both continues and differentiable. Its integral is given by $$\int_{s_3}^{s_3 + \gamma}Fd\lambda = -1$$  Moreover, $$\int_{s_3}^{s_3 + \gamma}\dot Fd\lambda = 0$$ Since the prefactors of both $F$ and $\dot F$ in the interval (in which $F$ and $\dot F$ are different then zero and do not change much), we can take them out of the integral, 
\begin{eqnarray}
\delta Q_c &=& \epsilon\int_{s_1}^{s_2}\left( \dot F\left( \Lambda-\frac{\dot w}{w\tau}  \right) +F\frac{ \Lambda \dot w}{ w}   \right) ds\nonumber\\
&=& \epsilon\left( \Lambda-\frac{\dot w}{w\tau}  \right)\int_{s_3}^{s_3 + \gamma} \dot Fds +\epsilon\frac{ \Lambda \dot w}{ w}  \int_{s_3}^{s_3 + \gamma}F ds\nonumber\\
&=& \epsilon\frac{ \Lambda \dot w}{w}   >0
\end{eqnarray}

This proves that there cannot be a protocol that attains the Carnot efficiency at finite time. 

Note that there are two scenarios in which the Carnot efficiency can nevertheless be attained: (i) There is no section $(s_1,s_2)$ in which the system is in contact with the coldest bath. As explained, the power in this case is zero. (ii) In the limit $\tau\rightarrow\infty$,  Eq.(\ref{Eq:Delta_Lambda}) gives finite $\delta\Lambda$ only if $\dot F(s)=0$ for all $s$, so our choice of $F$ is inconsistent, and the argument for no protocol with Carnot efficiency is invalid. This should be expected -- there are definitely protocols that saturate the Carnot bound in the $\tau\rightarrow\infty$ limit.
\end{document}